\documentclass[epj,referee]{svjour}
\usepackage{graphics}
\usepackage{epsfig}
\begin{document}

\title{}
\subtitle{Which mechanism underlies the water-like anomalies in core-softened potentials?}
\author{Alan Barros de Oliveira\inst{1} \and Paulo A. Netz \inst{2}\and Marcia C. Barbosa\inst{1}}                     

\institute{Instituto de F\'{\i}sica, Universidade Federal do Rio
Grande do Sul,
Caixa Postal 15051, 91501-970, Porto Alegre, Rio Grande do Sul, Brazil.  \and Instituto de Qu\'{\i}mica, Universidade Federal do Rio
Grande do Sul,
91501-970, Porto Alegre, Rio Grande do Sul, Brazil.}
\date{Received: \today / Revised version: date}

\abstract{Using molecular dynamics simulations 
we investigate the thermodynamic
of particles interacting  with  a continuous and a discrete versions of
a core-softened (CS) intermolecular
potential composed by a repulsive shoulder. Dynamic and structural
properties are also analyzed by the simulations.
We show that in the continuous version of the CS potential the density
at constant pressure has a maximum for a  certain temperature. Similarly the diffusion 
constant, $D$, at a constant temperature
has a maximum at a density $\rho_{\mathrm{max}}$ and a minimum at a density
$\rho_{\mathrm{min}}<\rho_{\mathrm{max}}$, and structural properties are also
anomalous. For the discrete CS potential none of these anomalies
are observed. The absence of anomalies in the discrete case and its
presence in the continuous CS potential are discussed
in the framework of the excess entropy.} 

\maketitle
%%%%%%%%%%%%%%%%%%%%%%%%%%%%
\section{Introduction}
\label{intro}
%%%%%%%%%%%%%%%%%%%%%%%%%%%%

Water is an anomalous substance in many respects. While most liquids contract upon cooling, 
for water the specific volume at ambient pressure starts to increase 
when cooled below $T=4 ^oC$ at atmospheric pressure \cite{Wa64}. This effect is called
density anomaly. Besides the density  anomaly, there are
more than sixty other anomalies
known for water \cite{URL}. The diffusivity  is one
of them. For normal liquids the diffusion coefficient,
$D$, decreases under compression. However, experimental results have shown that for water at
temperatures approximately below 10$^o$C the diffusion
coefficient increases under compression and has a
maximum \cite{An76}. The temperature of maximum density (TMD) line
inside which the density anomaly occurs, and
the line of maximum in diffusivity are located
in the same region of the pressure-temperature (P-T)
phase diagram of water. Simulations for water
also show thermodynamic and dynamic anomalies.
The simple point charged/extended (SPC/E) model for water
exhibits in the P-T phase diagram a TMD line.
The  diffusion coefficient has a maximum and a minimum
that define two lines at the P-T phase diagram,
the lines of maximum and minimum in the
diffusivity coefficient
\cite{Ne01,Er01,Mi06a}. Similarly to the experimental
results, the TMD and the lines of maximum and minimum
in the diffusion are located at the same
region at the P-T phase diagram
for the SPC/E model.
Errington and Debenedetti \cite{Er01} and Netz \emph{et al.} \cite{Ne01}
found, in SPC/E water,
that there exists a hierarchy between the density and diffusion anomalies as follows.
The diffusion anomaly region, inside which the
mobility of particles grow
as the density is increased, englobes the density anomaly region,
inside which the system
expands upon cooling at constant pressure. This observation 
is supported by experimental results \cite{An76}.

 Realistic simulations of water
\cite{St74,Be87,Jo00} have achieved a good accuracy in describing the
thermodynamic and dynamic anomalies of water. However, due to the high
number of microscopic details taken into account in these models, it
becomes difficult to discriminate what is essential to explain the
anomalies. On the other extreme, a number of isotropic models were
proposed as the simplest framework to understand the physics of
the  liquid state anomalies. From the
desire of constructing a simple two-body isotropic potential capable
of describing the complicated behavior present in water-like molecules,
a number of models in which single component systems of particles
interact via core-softened (CS) potentials have been proposed.
They possess a repulsive core that exhibits a region of
softening where the slope changes dramatically. This region can
be a shoulder or a ramp 
\cite{St98,Fr01,Ba04,Ol05,He05a,He05b,Ja98,Wi06,Ca03,Ol06a,Ol06b}. Ramp 
and continuous shoulder-like potentials exhibit
thermodynamic, dynamic, and structural anomalies
\cite{Ru06b,Ol06a,Ol06b}. However the square discontinuous shoulder 
shows no thermodynamic anomaly in three dimensions \cite{Fr01}.
All these potentials have in common the presence of two representative
repulsive scales in the potential, $\sigma$ and $\sigma_1$, where  the
closest scale, $\sigma<\sigma_1$, has 
the higher  potential energy,  $U(\sigma)>U(\sigma_1)$. 

One question that arises in this context is why a continuous shoulder
potential like the one described by de Oliveira \emph{et al.} \cite{Ol06a,Ol06b} 
has the anomalies while the square shoulder described by
Franzese et. al. \cite{Fr01} has no anomaly? 
In order to shade some light
in the reasons for the presence of
density anomalies in CS potentials, both the discontinuous shoulder 
potential and the continuous version are analyzed in
the framework of the excess-entropy-based formalism recently 
introduced by Errington
\emph{et al.}  \cite{Er06}. Within this approach the presence
of the density anomaly is related to the density dependence
of the excess entropy, $s_{\mathrm{ex}}$. 
We will follow this surmise and compute 
the excess entropy for both the discontinuous and the 
continuous shoulder potentials. From the  analysis of $s_{\mathrm{ex}}$
for the discontinuous model we are able to derive a simple 
argument for the presence or absence of anomalies in 
CS potentials.  

The outline of the paper is as follows.
We present the details of
the two models in Sec. \ref{sec:model}, details
of the simulations, and the P-T phase-diagram for
both models are presented in  Sec.  \ref{sec:sim}.
In Sec. \ref{sec:ex}  comparisons between the behavior
of the excess entropy of the continuous and discontinuous
models are made and from that a  condition
for the presence or absence of
anomalies in CS potentials is proposed. Conclusions end this
sections.

%%%%%%%%%%%%%%%%%%%%%%%%%%%%%%%%%%%%%%%%%%%%%%%%%%%%%%%%%%%
\section{The models} 
\label{sec:model}
%%%%%%%%%%%%%%%%%%%%%%%%%%%%%%%%%%%%%%%%%%%%%%%%%%%%%%%%%%%

\begin{figure}[htb]
\resizebox{0.6\columnwidth}{!}{\includegraphics[clip,width=7cm,height=5cm]{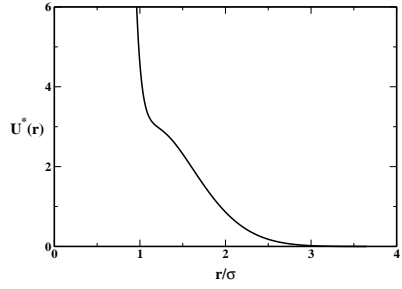}
}
\caption{Interaction potential from eq. (\ref{eq:potential1}) with
parameters $a=5,$ $r_{0}/\sigma=0.7$ and $c=1$, in reduced
units.  \label{cap:potential1}}
\end{figure}

The first model  we study here consists of a system of
$N$ particles  of diameter $\sigma,$ inside
a cubic box whose volume
is $V,$ resulting in a density number 
$\rho = N/V$ \cite{Ol06a} interacting with
a continuous shoulder potential given by

%%%%%%%%%%%%%%%%%%%%%%%%%%%%%%%%%%%%%%%%%%%%%%%%%%%%%%%%%%%
\begin{equation}
U^{*}(r)=4\left[\left(\frac{\sigma}{r}\right)^{12}-
\left(\frac{\sigma}{r}\right)^{6}\right]+
a\exp\left[-\frac{1}{c^{2}}\left(\frac{r-r_{0}}{\sigma}
\right)^{2}\right],
\label{eq:potential1}
\end{equation}
%%%%%%%%%%%%%%%%%%%%%%%%%%%%%%%%%%%%%%%%%%%%%%%%%%%%%%%%%%

\noindent where $U^{*}(r)=U(r)/\epsilon.$
The first term of Eq. (\ref{eq:potential1}) is a Lennard-Jones
potential of well depth $\epsilon$ and the second
term is a Gaussian centered on radius $r=r_{0}$ with height $a$
and width $c$.
In a previous publication we have studied 
this  model  with
setting $a=5,$ $r_{0}/\sigma=0.7$ and $c=1$
(see Fig. \ref{cap:potential1}) \cite{Ol06a}.
This potential has two  natural length scales: one close
to the
hard core, $\sigma,$ and another at
a further distance where the potential has its
lower value. This last length we call $\sigma_1$.
%%%%%%%%%%%%%%%%%%%%%%%%%%%%%%%%%%%%%%%
%%%%%%%%%%%%%%%%%%%%%%%%%%%%%%%%%%%%%%%

%%%%%%%%%%%%%%%%%%%%%%%%%%%%%%%%%%%%%%%
\begin{figure}[htb]
\resizebox{0.6\columnwidth}{!}{\includegraphics[clip,width=7cm,height=5cm]{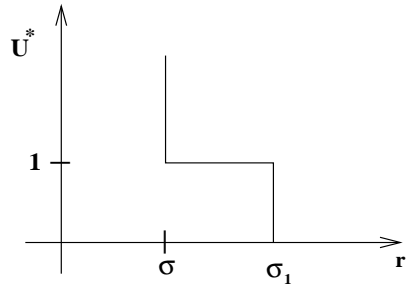}}
\caption{Interaction potential from eq. (\ref{eq:potential2}) with
parameter  $\sigma_1/\sigma=1.75$  in  reduced
units.  \label{cap:potential2}}
\end{figure}
%%%%%%%%%%%%%%%%%%%%%%%%%%%%%%%%%%%%%%%

The second model we study  here is a system of 
$N$ particles  of diameter $\sigma,$ inside
a cubic box whose volume
is $V,$ resulting in a density number 
$\rho = N/V$ but  interacting with a discontinuous
shoulder potential given by 
%%%%%%%%%%%%%%%%%%%%%%%%%%%%%%%%%%%%%%%%%%%%%%%%%%%%%%%%%%%
\begin{equation}
U^{*}(r)=\cases{\infty & $r<\sigma$ 
                   \cr 1 & $\sigma_1>r>\sigma$  \cr 0 &  $r>\sigma_1$  } \; ,
\label{eq:potential2}
\end{equation}
%%%%%%%%%%%%%%%%%%%%%%%%%%%%%%%%%%%%%%%%%%%%%%%%%%%%%%%%%%
\noindent where $U^{*}(r)=U(r)/\epsilon.$ This potential
has two natural length scales: the hard core distance, $\sigma,$ and
the outer core, $\sigma_1$. 
Here we analyze the case $\sigma_1/\sigma=1.75$ illustrated in 
Fig. \ref{cap:potential2}.
Here we use 
dimensionless pressure, $P^*,$ temperature, $T^*,$ and density, $\rho^*,$
that are  given in units of  $\sigma^3/\epsilon$, $k_{B} /\epsilon,$
and $\sigma^3,$ respectively and  $k_{B}$ stands for the 
Boltzmann constant.

%%%%%%%%%%%%%%%%%%%%%%%%%%%%%%%%%%%%%%%%%%%%%%%%%%%%%%%%%%%
\section{Details of simulations } 
\label{sec:sim}
%%%%%%%%%%%%%%%%%%%%%%%%%%%%%%%%%%%%%%%%%%%%%%%%%%%%%%%%%%%

For the continuous shoulder potential we
performed molecular dynamics simulations.
Details of the simulation 
can be found in ref \cite{Ol06a}.
%%%%%%%%%%%%%%%%%%%%%%%%%%%%%%%%%%%%%%%%%%%%
\begin{figure}[htb]
\resizebox{0.55\columnwidth}{!}{\includegraphics[clip,width=7cm,height=5cm]{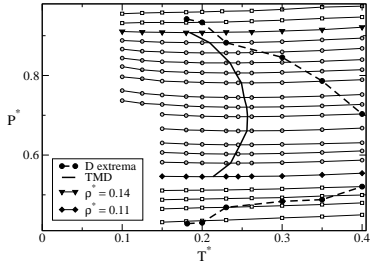}}
\caption{Pressure-temperature phase-diagram obtained for the 
continuous shoulder potential. From
bottom to top
$\rho^* =$ 0.100, 0.103, 0.105, 0.107
0.110, 0.113, 0.115, 0.117, 0.120, 0.123,
0.125, 0.127, 0.130, 0.132, 0.134, 0.136,
0.138, 0.140, 0.142, and 0.144
are shown.
The solid line illustrates the TMD and the dashed 
lines show the boundary of the diffusivity extrema.
 \label{cap:p-t}}
\end{figure}
%%%%%%%%%%%%%%%%%%%%%%%%%%%%%%%%%%%%%%%%%%%%
%%%%%%%%%%%%%%%%%%%%%%%%%%%%%%%%%%%%%%%%%%%%
\begin{figure}[htb]
\resizebox{0.55\columnwidth}{!}{\includegraphics[clip,width=7cm,height=5cm]{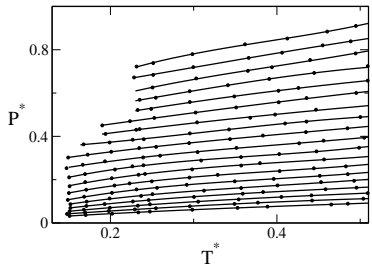}}
\caption{Pressure-temperature phase-diagram for the discontinuous
shoulder potential. From
bottom to top
$\rho^* =$ 0.08, 0.09, 0.10, 0.11, 0.12, 0.13, 0.14,
0.15, 0.16, 0.17, 0.18, 0.19,
0.20, 0.21, 0.22, 0.23, 0.24, 0.25, and 0.26 are shown.
 \label{cap:p-t2}}
\end{figure}
%%%%%%%%%%%%%%%%%%%%%%%%%%%%%%%%%%%%%%%%%%%%
Figure (\ref{cap:p-t}) shows the P-T phase-diagram we have
obtained in a previous publication \cite{Ol06a}. The isochores have minima that define the 
temperature of maximum density. The TMD line encloses the region of density
(and entropy) anomaly.
We also have studied the mobility associated with the potential described in
Eq.~(\ref{eq:potential1}) \cite{Ol06a}.  The
diffusion was also calculated using the the
mean-square displacement averaged over different initial times.
The behavior of $D$ as a function of $\rho^*$ goes as follows.
At low temperatures, the behavior
is similar to the behavior found in SPC/E supercooled
water \cite{Ne01}. The diffusivity increases as the density is lowered, reaches a
maximum at $\rho_{D{\rm max}}$ (and $P_{D{\rm max}}$) and decreases
until it reaches a minimum at $\rho_{D{\rm min}}$ (and $P_{D{\rm min}}$).
The region in the P-T plane where there is an anomalous behavior in
the diffusion is bounded by $(T_{D{\rm min}},$ $P_{D{\rm min}})$ and
$(T_{D{\rm max}},$ $P_{D{\rm max}})$ and their location is shown
in Fig.~ (\ref{cap:p-t}).
The region of diffusion anomalies $(T_{D{\rm max}}$, 
$P_{D{\rm max}})$ and $(T_{D{\rm min}},$ $P_{D{\rm min}})$ 
lies outside the region of density anomalies like in SPC/E water \cite{Ne01}.

In order to simulate the discrete potential showed in Fig. (\ref{cap:potential2})  we used the collision
driven molecular dynamics techniques \cite{Al59}. 500 particles were put into a simulation box with periodic 
boundary conditions and the rescaling velocities scheme was used for every 1000 collisions until 
equilibration time to achieve the desired temperature. After thermalization particles were 
allowed to move under microcanonical ensemble. 
In the collision driven molecular dynamics simulation, kinetic energy has to be rigorously conserved. 
Hence, no special mechanism is necessary in order to simulate a desired temperature and the 
NVE ensemble becomes the natural choice. 
The equilibration and production times in reduced units were 350 and 650 respectively. 
Figure (\ref{cap:p-t2}) shows the P-T phase-diagram for the  discontinuous shoulder potential. 
The isochores for
different temperature and pressures  show  no minima so no density anomaly is present.

%%%%%%%%%%%%%%%%%%%%%%%%%%%%%%%%%%%
\section{Excess entropy and anomalies}
\label{sec:ex}
%%%%%%%%%%%%%%%%%%%%%%%%%%%%%%%%%%%
Why the discontinuous shoulder 
potential  has no water-like anomalies and
its continuous counterpart has all of them? We
can gain some understanding about 
 that by analyzing the density
dependence of the excess entropy.  
Errington
\emph{et al.} have shown that the 
density anomaly is given by the condition  
$\Sigma_{\mathrm{ex}}= \left(\partial s_{\mathrm{ex}} /\partial \ln \rho\right)_T>1$
\cite{Er06}. Here $s_{\mathrm{ex}}$ is the excess entropy and is approximated  by
its two-body contribution,

$$
s_{2}=-2\pi\rho\int\left[g(r)\ln g(r)-g(r)+1\right]r^2 dr \;.
$$

The radial
distribution  function, $g(r)$,  is proportional to the
probability to find a particle at a distance $r$ to another particle
placed at the origin.
Errington \emph{et al.} \cite{Er06} have also suggested that the diffusion anomaly
can be predicted by using the empirical Rosenfeld's parameterization \cite{Ro99}. They found
the condition  $\Sigma_{2} > 0.42$ for a diffusion anomalous behaviour.  They also
claim that $\Sigma_{2} > 0$ 
is a good estimative for determining the region where structural anomaly occurs \cite{Er06}.

In order to understand the differences between the continuous and
the discontinuous shoulder potentials we test the excess entropy 
criteria described 
above in both potentials. 
The radial distribution functions for different
temperatures and densities for both potentials were then
obtained by the molecular dynamic simulation method described
in Sec. \ref{sec:sim}.

\begin{figure}
\includegraphics[%
  clip,
  scale=0.25]{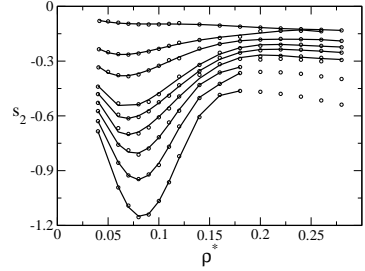}
\caption{Pair contribution of excess entropy, $s_2$, against density
for $T^* =$ 0.25, 0.30, 0.35, 0.40, 0.45, 0.50, 0.70, 1.00, and 3.00  from bottom to top
for the continuous shoulder model (Fig.(\ref{cap:potential1}) . Circles
are simulated data and lines are fifth order polynomial fit from data.  \label{cap:sex-potential1}}
\end{figure}

\begin{figure}
\includegraphics[%
  clip,
  scale=0.25]{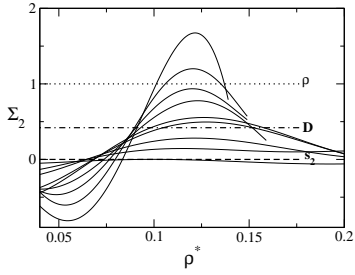}
\caption {$\Sigma_{2}= \left(\partial s_2 / \partial \ln \rho \right)_T$
versus density.
Following the isochore $\rho^* = 0.10$, from
top to bottom, temperatures $T^* =$
0.25, 0.30, 0.35, 0.40, 0.45, 0.50, 0.70, 1.00, and 3.00 are shown
for the continuous shoulder
potential. \label{cap:dsex-potential1}}
\end{figure}

Figure (\ref{cap:sex-potential1}) illustrates the two-body contribution
of excess entropy for the
continuous potential given by Eq. (\ref{eq:potential1}). $s_2$
is negative and its slope changes from positive to negative what indicates
the presence of structural anomaly. Figure (\ref{cap:dsex-potential1})
shows the behavior of $\Sigma_{2}$ with density for a fixed temperature
for the continuous model.
The horizontal lines at $\Sigma_{2}^t =$ 0, 0.42, and 1 indicate the threshold
beyond which there are structural, diffusion, and density anomalies respectively. 
In accordance with Fig. (\ref{cap:p-t}) the density anomalous 
region shown in Fig. (\ref{cap:dsex-potential1}) occurs in an interval
of density smaller than the interval where the diffusion is anomalous.
The two-body excess entropy also show the presence of structural anomaly
what corroborates results of simulations of structural parameters \cite{Ol06b}.

\begin{figure}
\includegraphics[%
  clip,
  scale=0.25]{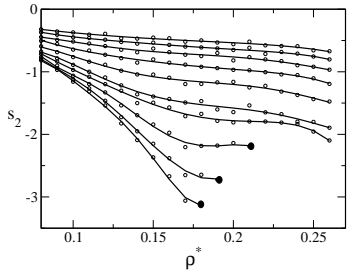}
\caption{Pair contribution of excess entropy, $s_2,$ against density
for $T^* =$ 0.15, 0.17, 0.20, 0.23, 0.25, 0.30, 0.35, 0.40, 0.45, and 0.50 
from bottom to top for the 
discontinuous shoulder model (Fig.(\ref{cap:potential2}). Open circles
are simulated data and lines are fifth order polynomial fit from data. Solid circles show
the limit of crystalization.   \label{cap:sex-potential2}}
\end{figure}

\begin{figure}
\includegraphics[%
  clip,
  scale=0.25]{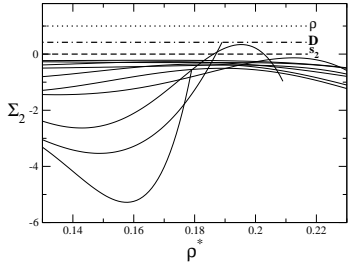}
\caption{ $\Sigma_{2}= \left(\partial s_2 / \partial \ln \rho \right)_T$
versus density for the discontinuous shoulder  potential. Following the isochore
$\rho^* =0.13,$ from bottom to top, $T^* =$ 0.15, 0.17, 0.20, 0.23, 0.25, 0.30, 0.35, 0.40, 0.45, and 0.5 
are shown.  \label{cap:dsex-potential2}}
\end{figure}

Figure (\ref{cap:sex-potential2}) illustrates the two-body excess entropy of the
discontinuous potential given by Eq. (\ref{eq:potential2}). $s_2$
is negative but its slope is for almost all temperatures
and densities negative. The filled circles show the 
region where the system crystallization occur.  
Figure (\ref{cap:dsex-potential2}) shows the behavior of $\Sigma_{2}$ with density for 
fixed temperatures for the discontinuous model.
The line  $\Sigma_{2}^t=1$ is never crossed so no
density anomaly should be expected  what is in good
agreement with the Fig. (\ref{cap:p-t2}). The line $\Sigma_{2}^t=0.42$
is also never crossed what indicates that diffusion
anomaly is not expected in the discontinuous model. This 
is also in agreement with results for similar step potentials
where no diffusion anomaly is found for large steps  \cite{Ne06}.
The line $\Sigma_{2}=0$ is crossed for temperatures
$T^* =$ 0.17 and 0.2 what would suggest the presence of
structural anomaly. However this has to be taken with a grain of salt 
since  Errington \emph{et al.}  \cite{Er06} demonstrated that
$s_2$ overestimates the region of anomalies. In this sense, 
a detailed study of translational and orientational order parameters
\cite{Ol06b} is necessary prior to any affirmative on the presence of
structural anomaly for this discontinuous shoulder.

\begin{figure}
\includegraphics[%
  clip,
  scale=0.25]{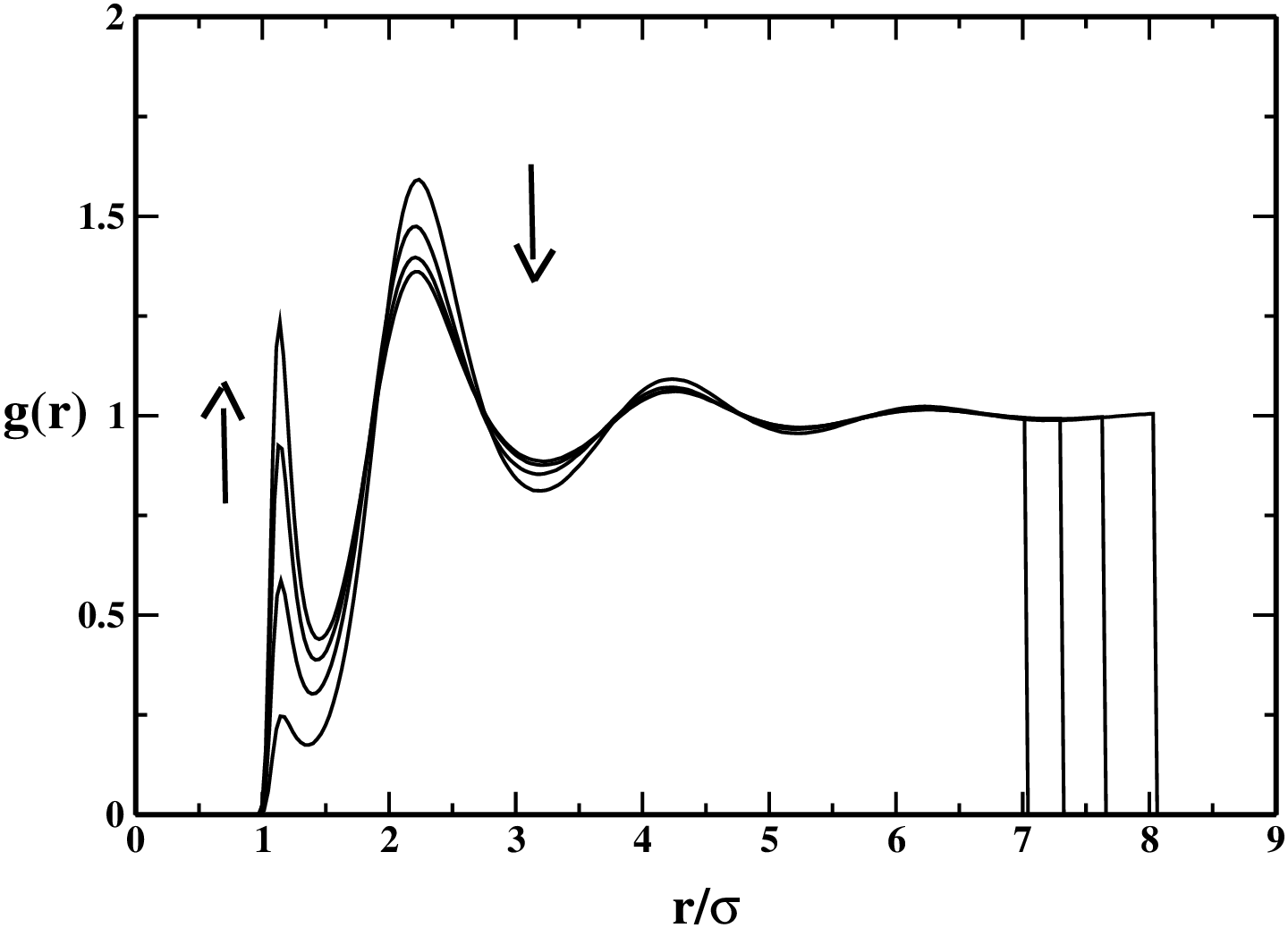}
\caption{Radial distribution function for the potential
given by Eq. (\ref{eq:potential1}) for $T^*=0.25$ and
densities $\rho^* =$ 0.12, 0.14, 0.16, and 0.18.
Arrows indicate the direction of
increasing $\rho^{*}.$   \label{cap:gr-potential1}}
\end{figure}

\begin{figure}
\includegraphics[%
  clip,
  scale=0.25]{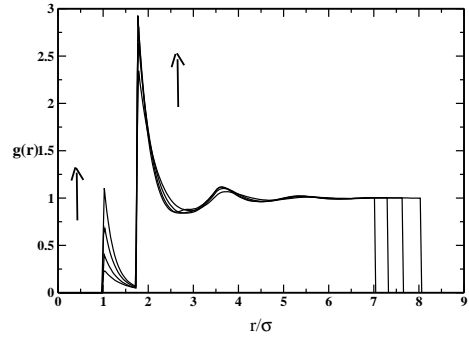}
\caption{Radial distribution function for the potential
given by Eq. (\ref{eq:potential2}) for $T^*=0.25$ and
densities  $\rho^* =$ 0.12, 0.14, 0.16, and 0.18.
Arrows indicate the direction of
increasing $\rho^{*}.$   \label{cap:gr-potential2}}
\end{figure}

Even thought the excess entropy criteria  is quite useful for predicting if
the anomalies would be present for a certain potential it does not
provide an easy and intuitive method for understanding why
the continuous shoulder has anomalies and the discontinuous one
does not have. 
In principle, both potential exhibit similar characteristics. 
Both potentials have two repulsive scales and
consequently the 
radial distribution function in both cases has two peaks, one
close to the hard core and another close to the distance $r=\sigma_1$ as illustrated
in Fig. (\ref{cap:gr-potential1}) for the continuous potential
and in Fig. (\ref{cap:gr-potential2}) for the discontinuous potential.
A closer look at the radial distribution reveals an important
difference between the two cases. For the continuous potential
for densities and temperatures in the  region where anomalies occur
$g(r)$ grows with density for $r/\sigma \approx 1$ and decreases 
with increasing density for $r\approx \sigma_1$ [see Fig. (\ref{cap:gr-potential1}) for example]. 
For the discontinuous potential the $g(r)$ increases with density both at the hard core
and at $r=\sigma_1$ for any temperature and density [see Fig. (\ref{cap:gr-potential2}) for example].  
Notice that the radial distribution function  in both cases has significative
changes with the change in density at the two natural scales.

Finally we propose  that a two scale potential has 
anomalies for some temperature and 
densities if  
$\partial g(r)/\partial\rho>0$  for $r\approx \sigma$  
and $\partial g(r)/\partial\rho<0 $  for $r\approx \sigma_1$.
If this requirements would not be fulfill for
any temperature and density no anomaly would be present.
This proposition is based in the physical picture that
within the anomalous region for a fixed temperature an increase
is density implies an increase in the number of particles 
close to the hard core. This particles move from 
the distance $\sigma_1$ to $\sigma$. For a discontinuous 
potential this requires an activation energy of $U=\epsilon$ while
for the continuous case it can be done continuously.
As it was shown by Netz \emph{et al.} \cite{Ne06} for
the steps potential, anomalies would only be observed
if the discontinuity in $U^*$ would be below a
 certain threshold.

Now we shall test if this simple hypothesis is in agreement 
with Errington \emph{et al.}  criteria. First we compute $\Sigma_{2}$ as
%%%%%%%%%%%%%%%%%%%%%%%%%%%%%%%%%%%
\begin{eqnarray*}
\label{eq:sigma}
\Sigma_{2} & = & \left(\frac{\partial s_{\mathrm{{ex}}}}{\partial\ln\rho}\right)_{T}\\
 & = & s_{2}-2\pi\rho^{2}\int\ln g(r)\frac{\partial g(r)}{\partial\rho}r^{2}dr\end{eqnarray*}
%%%%%%%%%%%%%%%%%%%%%%%%%%%%%%%%%%%  
In this expression, the first term is always negative for
all two scale potentials. The integral in the second
term, as we have observed above,  is dominated by the the values at the 
two scales, $\sigma$ and $\sigma_1$. 

%%%%%%%%%%%%%%%%%%%%%%%%%%%%%%%%%%%%%%%

%%%%%%%%%%%%%%%%%%%%%%%%%%%%%%%%%%%%%%%

%%%%%%%%%%%%%%%%%%%%%%%%%%%%%%%%%%%%%%%

%%%%%%%%%%%%%%%%%%%%%%%%%%%%%%%%%%%%%%%

%%%%%%%%%%%%%%%%%%%%%%%%%%%%%%%%%%%%%%%
%%%%%%%%%%%%%%%%%%%%%%%%%%%%%%%%%%%%%%%

%%%%%%%%%%%%%%%%%%%%%%%%%%%%%%%%%%%%%%%
%%%%%%%%%%%%%%%%%%%%%%%%%%%%%%%%%%%%%%%

%%%%%%%%%%%%%%%%%%%%%%%%%%%%%%%%%%%%%%%
%%%%%%%%%%%%%%%%%%%%%%%%%%%%%%%%%%%%%%%
%%%%%%%%%%%%%%%%%%%%%%%%%%%%%%%%%%%%%%%
%%%%%%%%%%%%%%%%%%%%%%%%%%%%%%%%%%%%%%%

For the continuous potential in the region where the anomaly
is present $\ln g(r\approx \sigma)<1$ 
while   $\ln g(r\approx\sigma_1)>1$. 
Also $\partial g(\sigma)/\partial \rho>0$ 
while $\partial g(\sigma_1)/\partial \rho<0$. Consequently 
the second term in Eq. (\ref{eq:sigma}) is positive. This 
allows for a zero or positive values of $\Sigma_2$ for
appropriated densities and temperatures. Therefore
our criteria is in accord with
Errington
\emph{et al.} criteria.

In resume, in this paper we have calculated the 
excess entropy and its derivative for both
continuous and discontinuous shoulder potential. For
the continuous case, using 
the Errington
\emph{et al.} criteria indicates that this potential
has density, diffusion and structural anomalies as we
have shown in previous publications. For the 
discontinuous potential the criteria indicates that
no thermodynamic and dynamic anomalies are present
and its not conclusive for structural anomalies. 
Direct calculations of the P-T temperature
phase-diagram confirms the excess of entropy
prediction. On basis of these results we propose
a  criteria
for predicting if a two scale potential has or
not anomalies. Our criteria provides a simple
picture for the anomalies not being observed 
in the discontinuous shoulder potential.

We thank for financial support of
the Brazilian science agencies CNPq, CAPES, and FINEP.
One of the authors (A.B.O.) is indebted to
Jeetain Mittal of NIH 
for his valuable discussions
on collision driven molecular dynamics 
techniques.

\bibliographystyle{epj}
\bibliography{Biblioteca}

\begin{thebibliography}{25}

\bibitem{Wa64}
R.~Waler, \emph{Essays of natural experiments} (Johnson Reprint, New York,
  1964)

\bibitem{URL}
M.~Chaplin, \emph{Sixty-three anomalies of water},
  http:$//www.lsbu.ac.uk/water/anmlies.html$ (2006)

\bibitem{An76}
C.A. Angell, E.D. Finch, P.~Bach, J. Chem. Phys. \textbf{65}, 3065 (1976)

\bibitem{Ne01}
P.A. Netz, F.W. Starr, H.E. Stanley, M.C. Barbosa, J. Chem. Phys. \textbf{115},
  344 (2001)

\bibitem{Er01}
J.R. Errington, P.D. Debenedetti, Nature (London) \textbf{409}, 318 (2001)

\bibitem{Mi06a}
J.~Mittal, J.R. Errington, T.M. Truskett, J. Phys. Chem. B \textbf{110}, 18147
  (2006)

\bibitem{St74}
F.H. Stillinger, A.~Rahman, J. Chem. Phys. \textbf{60}, 1545 (1974)

\bibitem{Be87}
H.J.C. Berendsen, J.R. Grigera, T.P. Straatsma, J. Chem. Phys. \textbf{91},
  6269 (1987)

\bibitem{Jo00}
M.W. Mohoney, W.L. Jorgensen, J. Chem. Phys. \textbf{112}, 8910 (2000)

\bibitem{St98}
M.R. Sdr-Lahijany, A.~Scala, S.V. Buldyrev, H.E. Stanley, Phys. Rev. Lett.
  \textbf{81}, 4895 (1998)

\bibitem{Fr01}
G.~Franzese, G.~Malescio, A.~Skibinsky, S.V. Buldyrev, H.E. Stanley, Nature
  (London) \textbf{409}, 692 (2001)

\bibitem{Ba04}
A.~Balladares, M.C. Barbosa, J. Phys.: Cond. Matter \textbf{16}, 8811 (2004)

\bibitem{Ol05}
A.B. de~Oliveira, M.C. Barbosa, J. Phys.: Cond. Matter \textbf{17}, 399 (2005)

\bibitem{He05a}
V.B. Henriques, M.C. Barbosa, Phys. Rev. E \textbf{71}, 031504 (2005)

\bibitem{He05b}
V.B. Henriques, N.~Guissoni, M.A. Barbosa, M.~Thielo, M.C. Barbosa, Mol. Phys.
  \textbf{103}, 3001 (2005)

\bibitem{Ja98}
E.A. Jagla, Phys. Rev. E \textbf{58}, 1478 (1998)

\bibitem{Wi06}
H.M. Gibson, N.B. Wilding, Phys. Rev. E \textbf{73}, 061507 (2006)

\bibitem{Ca03}
P.~Camp, Phys. Rev. E \textbf{68}, 061506 (2003)

\bibitem{Ol06a}
A.B. de~Oliveira, P.A. Netz, T.~Colla, M.C. Barbosa, J. Chem. Phys.
  \textbf{124}, 084505 (2006)

\bibitem{Ol06b}
A.B. de~Oliveira, P.A. Netz, T.~Colla, M.C. Barbosa, J. Chem. Phys.
  \textbf{125}, 124503 (2006)

\bibitem{Ru06b}
R.~Sharma, S.N. Chakraborty, C.~Chakravarty, J. Chem. Phys. \textbf{125},
  204501 (2006)

\bibitem{Er06}
J.R. Errington, T.M. Truskett, J.~Mittal, J. Chem. Phys. \textbf{125}, 244502
  (2006)

\bibitem{Al59}
B.J. Alder, T.E. Wainwright, J. Chem. Phys. \textbf{31}, 459 (1959)

\bibitem{Ro99}
Y.~Rosenfeld, J. Phys.: Condens. Matter \textbf{11}, 5415 (1999)

\bibitem{Ne06}
P.A. Netz, S.~Buldyrev, M.C. Barbosa, H.E. Stanley, Physical Review E
  \textbf{73}, 061504 (2006)

\end{thebibliography}

\end{document}